\documentstyle[11pt]{article}
\newcommand{\beq}{\begin{equation}}
\newcommand{\eeq}{\end{equation}}
\newcommand{\bea}{\begin{eqnarray}}
\newcommand{\eea}{\end{eqnarray}}

\def\lm{\lambda}
\def\om{\omega}
\def\omo{\omega_{1}}
\def\omt{\omega_{2}}
\def\lno{\langle n^{(0)} | }
\def\rno{ | n^{(0)} \rangle }
\def\lmo{\langle m^{(0)} |}
\def\rmo{ | m^{(0)} \rangle }
\def\bH{\bar{H}}
\def\bHo{\bar{H}^{(1)}}
\def\bHt{\bar{H}^{(2)}}
\def\Ho{H^{(1)}}
\def\Ht{H^{(2)}}
\def\wo{w^{(1)}}
\def\wt{w^{(2)}}
\def\bx{{\bf x}}
\def\bbx{\bar{{\bf x}}}
\def\bz{\bar{z}}
\def\zo{z_{1}}
\def\zt{z_{2}}
\def\bzo{\bar{z}_{1}}
\def\bzt{\bar{z}_{2}}
\def\var{\varepsilon}
\def\bc{\bar{a}}
\def\pr{\partial}
\begin{document}
\renewcommand{\thefootnote}{\fnsymbol{footnote}}
                                        \begin{titlepage}
\begin{flushright}
TECHNION-PHYS-95-1 \\
hep-ph/9503240 \\
March 1995
\end{flushright}
\vskip1.8cm
\begin{center}
{\LARGE
Quantum KAM Technique and Yang-Mills Quantum Mechanics
            \\ }
\vskip1.5cm
 {\Large Igor~Halperin}
 \\
\vskip0.2cm
       Technion - Israel Institute of Technology   \\
       Department of Physics  \\
       Haifa, 32000,  Israel \\
{\small e-mail address: higor@techunix.technion.ac.il}\\

\vskip1.5cm

{\Large Abstract:\\}
\parbox[t]{\textwidth}{
We study a quantum analogue of the iterative perturbation theory by
Kolmogorov used in the proof of the Kolmogorov-Arnold-Moser (KAM)
theorem. The method is based on sequent canonical transformations with a
"running" coupling constant $ \lm , \lm^{2} , \lm^{4} $, etc.
The proposed scheme,
as its classical predecessor, is "superconvergent" in the sense that
after the nth step, a theory is solved to the accuracy  of order
$ \lm^{2^{n-1}} $ . It is shown
that the Kolmogorov technique corresponds to an infinite resummation
of the usual perturbative series. The corresponding expansion is
convergent for the quantum anharmonic oscillator due to the fact that
it turns out to be identical to the Pade series.
The method is easily generalizable to many-dimensional cases.
The Kolmogorov technique is further
applied to a non-perturbative treatment
of Yang-Mills quantum mechanics. A controllable expansion for the
wave function near the origin is constructed. For large fields,
we build an asymptotic adiabatic expansion in inverse powers of the
field. This asymptotic solution contains arbitrary constants which
are not fixed by the boundary conditions at infinity. To find them,
we approximately match the two expansions in an intermediate region
.We also discuss some analogies between this problem and the method of QCD
sum rules.  }
\vspace{1.0cm}
\end{center}
                                                \end{titlepage}

\section{Introduction}

The remarkable progress made in a last few decades in the study of
dynamical systems in classical mechanics has stimulated a
renewed interest and revision of fundamentals of quantum mechanics.
In particular, a number of efforts has been undertaken to
elucidate possible manifestations of a (semi-) quantum chaos which,
due to the correspondence principle, is expected to persist in this
or that form in quantum mechanics, and to translate the concepts of
non-linear classical mechanics to the quantum language ( see e.g.
\cite{Gut} for a review ). One may hope, at the same time, that
powerful calculation schemes of classical mechanics, being respectively
transformed, could yield new methods supplementing more traditional
tools of quantum physics such as the perturbation theory.
\newline
  One such an approach is the
                                          Kolmogorov superconvergent
method of classical mechanics \cite{Kol} which has been the working
instrument in the proof of the celebrated KAM theorem \cite{ArMos} .
In a shortened form, the main Kolmogorov's idea can be introduced
as follows. Imagine we have added some non-integrable perturbation
$ \varepsilon V_{1} (I, \theta ) $ to a integrable Hamiltonian
$ H_{0} ( I) $ which is the function of the action variables only.
It has been known for a long time \cite{Bir} that successive
canonical transformations $ (I, \theta ) \rightarrow ( I_{1} ,
\theta_{1} )   \rightarrow ...
\rightarrow ( I_{n} , \theta_{n} ) $ are able to lower the formal
order of the perturbation by one order of the small parameter
$ \varepsilon $ at every step $ \varepsilon V_{1} \rightarrow
 \varepsilon^{2} V_{2} \rightarrow
... \rightarrow  \varepsilon^{n} V_{n} $ .
However, this procedure generally results in a diverging
asymptotic series applicable only at fairly short time
intervals. The divergence originates in a possibility for
an existence of rational ( and close to rational ) resonant
relations among the frequencies $ \omega_{i} $ of the unperturbed
motion $ \sum_{i} m_{i} \omega_{i} = 0 $ with a set of the integer
$m_{i}$'s . This is the so-called small denominators problem ( we
refer the interested reader to the classical textbooks \cite{Arn},
\cite{Lic} and the excellent review paper \cite{Chir} on the whole
subject of the classical perturbation theory and the KAM theorem ).
As any irrational number can be approximated to an arbitrary
accuracy by rational numbers, one can readily imagine the scale
of difficulties to be met with. It was Kolmogorov \cite{Kol} who
has managed to show for the first time that successive canonical
transformations may be chosen such that the order of the
perturbation is increased by the {\it square} of the
proceeding one for each step : $ \varepsilon V_{1} \rightarrow
 \varepsilon^{2} V_{2} \rightarrow \varepsilon^{4} V_{3}
 ... \rightarrow  \varepsilon^{2^{n-1}} V_{n}
$ . This phenomenon of {\it superconvergence} occurs as a result
of a clever redefinition of what is to be named "unperturbed" and
"perturbating" Hamiltonian at each iteration step. This idea actually
dates back to the well known Newton method for finding the root
of the equation f(x) = 0 . Instead of expanding the function into
the Taylor series around a suspected point $ x_{0} $ , the Newton
method suggests a simple first order iterative procedure of the
form $ x_{n+1} = x_{n} - \frac{f(x_{n})}{f'(x_{n})} $ . Thus, the
value about which the expansion is made is moved closer to the
true root at each iteration, that results in the superconvergence
of the Newton method ( more on the analogy between the Newton
method and the Kolmogorov approach can be found in \cite{Lic} ).
It turns out that owing to the superconvergence, the new
perturbation theory is able to overcome the effect of near-to-
resonance small denominators. Consequently, we end up with the
convergent procedure, provided the initial conditions are
adjusted at each step such that not to get just on a resonance
surface \cite{ArMos},\cite{Arn},\cite{Lic}, \cite{Chir}.
It has been shown \cite{ArMos}
how to estimate from below the smallness of the resonant
denominators that represent a set of frequencies of motion of a
measure zero. The KAM theorem \cite{Kol},\cite{ArMos} states that
most non-resonant invariant tori are only slightly deformed by
small smooth Hamiltonian perturbations and "form the majority
in the sense that the measure of the complement of their union
is small when the perturbation is small" \cite{Arn}.
\newline
  In this paper we study a quantum analogue of the Kolmogorov
technique. We find that, besides of the obvious aesthetic attractiveness
(the superconvergence in the above sense), the novel scheme possesses
other interesting  properties.
In particular, it will be shown that the constructed
superconvergent procedure corresponds to an infinite
resummation of the usual perturbative series and in some (non-trivial)
cases yields a {\it convergent} quantum iterative scheme. Another
observation is that the superconvergent method can be of use in cases
when the usual perturbation theory fails. To exemplify this
property, we work out so-called Yang-Mills quantum mechanics
\cite{Sav}.
Besides being a model for an infra-red behavior of the YM fields
\cite{Bar} ,
this system is of particular interest due to the fact that its
classical counterpart possesses strong chaotic properties \cite{Cha} .
Note in this respect that the most popular nowadays approach to the
quasiclassical quantization of classically chaotic systems, based
on a summation of periodic orbit \cite{Gut} , may be not the best
way in view of its formidable complexity (the number of periodic
orbits grows exponentially with the energy in a system with
hard chaos).
 Sometimes it can be expected
that the corresponding purely quantum system is much easier to solve.
Then, given a quantum solution, the quasiclassical limit can be followed
up. Finally, we mention that though our presentation is restricted by
quantum mechanical examples, the formalism may hopefully be extended to
the field theory.
\newline
The presentation is organized as follows.
In Sect.2 we solve the standard textbook problem of building
the quantum mechanical perturbation theory in a somewhat unusual
manner which elucidates the quantum superconvergence phenomenon.
The practical inconvenience of this approach turns out to be similar
to that of the classical perturbation theory based on canonical
transformations (see below). Then in Sect.3 we present an equivalent
but more convenient "non-coordinate" approach based on the so-called
Lie transformation technique (see e.g \cite{Lic} ). In this form, the
method admits ( at least, formally ) a generalization to
many-dimensional cases. Its additional advantage is a formal
equivalence between the classical and quantum perturbation theory.
The Kolmogorov technique within the Lie transforms is further
outlined. The proposed scheme is applied in Sect.4 to the classical
anharmonic oscillator and the reconstruction of the perturbative
series is demonstrated. The corresponding problem for the quantum
anharmonic oscillator is studied in Sect.5 . The obtained series
is shown to be equivalent to the well known Pade series,
thus providing convergent expansions for both small and strong
couplings. We check by direct calculations that a re-expansion of
the Kolmogorov series back into the standard power series reproduces
the known results for the ground state energy.
 Finally, we apply in Sect.6 the superconvergent
method to physically interesting YM quantum  mechanics.
 Avoiding the small denominators to lowest orders is
demonstrated. We show that the modified perturbation
theory yields a controllable approximation to the behavior of the
wave function near the origin. In a large fields region, we make
use of another approximation and obtain an asymptotic adiabatic
expansion in inverse powers of a field. This solution contains
arbitrary constants which are to be determined from a matching
with the small distance expansion. We make the approximate
matching in two ways. The first one is essentially the least
square matching over a "stability region" similar to the procedure
used in the QCD sum rules method \cite{Shif}. The second approach
is a matching of the wave functions and their derivatives in a
distribution sense (see below). Both methods yield similar results.
The ground state energy is calculated with an accuracy of order
3.5 \% while the accuracy of matching the wave functions is
worse, of order  15 \% . Sect.7 contains a brief summary and several
concluding remarks.

\section{Superconvergent perturbation theory}

Let us suppose we know the solution to the unperturbed problem
\beq
H_{0} \phi_{n}^{(0)} = E_{n}^{(0)} \phi_{n}^{(0)}
\eeq
and we try to solve the perturbed problem
\beq
(H_{0} + \lm W_{0}) \Psi_{n} = E_{n} \Psi_{n}
\eeq
We would like to solve this standard textbook's problem by a nonstandard
method following as close as possible the Kolmogorov's idea of
superconvergence (see Sect.3 for the corresponding classical treatment).
To this end, let us look for a solution in the form of a unitary
transform
\beq
\Psi_{n} = ( e^{ - \lm S_{1}} \phi^{(1)} )_{n} = \phi_{n}^{(1)} -
   \lm \sum_{m \neq n} \langle n^{(1)} | S_{1} | m^{(1)} \rangle
       \phi_{m}^{(1)} + \ldots
\eeq
where $ S_{1} $ is an anti-hermitean operator $ S_{1}^{+} = - S_{1} $.
We try to choose $ S_{1} $ in such way that the resulting equation
for $ \phi^{(1)} $ will be in some sense simpler than the initial
equation (2). We obtain
\beq
e^{ \lm S_{1}} ( H_{0} + \lm W_{0} ) e^{ - \lm S_{1}}
  \phi^{(1)} = E \phi^{(1)}
\eeq
or
\beq
( H_{0} + \lm ( W_{0} + [ S_{1} , H_{0} ] ) + \lm^{2} W_{2} +
\lm^{3} W_{3} + \lm^{4} W_{4} + O( \lm^{5} ) ) \phi^{(1)} =
E \phi^{(1)}
\eeq
where
\bea
W_{2} &=& [S_{1} , W_{0} ] + \frac{1}{2} [ S_{1} , [ S_{1} , H_{0} ]
      ]     \nonumber   \\
W_{3} &=& \frac{1}{2}   [S_{1},[S_{1},W_{0}]] + \frac{1}{3!} [S_{1},
[S_{1},[S_{1}, H_{0}]]] \\
W_{4} &=& \frac{1}{3!} [S_{1},[S_{1},[S_{1},W_{0}]]] + \frac{1}{4!}
[S_{1},[S_{1},[S_{1},[S_{1},H_{0}]]]]  \nonumber
\eea
Let us require that the operator $ W_{0} + [S_{1},H_{0}] $ has only
diagonal matrix elements in the basis of the eigenfunctions of the
operator $ H_{0} $ :
\bea
\lno W_{0} + [S_{1},H_{0}] \rmo &=& 0 \; , \; n \neq m \nonumber \\
\lno W_{0} + [S_{1},H_{0}] \rno &=& \lno W_{0} \rno
\eea
It follows from (7) that
\beq
\lno S_{1} \rmo = \frac{ \lno W_{0} \rmo }{ E_{n}^{(0)} - E_{m}^{(0)}}
           \; , \; n \neq m
\eeq
If we omit the terms $ O( \lm^{2}) $ and higher, we thus obtain the
usual formulas of the first order perturbation theory. Indeed, to
this accuracy the exact solution to the equation (6) is just the
unperturbed initial wave function : the equation
\beq
H_{1} \phi_{n} \equiv ( H_{0} + \lm ( W_{0} + [S_{1},H_{0}])
 \phi_{n} = E_{n} \phi_{n}
\eeq
has the solution
\beq
\phi_{n} = \phi_{n}^{(0)} \; , \; E_{n} = E_{n}^{(0)} +
\lm \lno W_{0} \rno
\eeq
and thus the solution to our problem to the given accuracy is
\beq
\Psi_{n}^{(1)} = \phi_{n}^{(0)} + \lm \sum_{m \neq n} \frac{ \lmo
W_{0} \rno }{ E_{n}^{(0)} - E_{m}^{(0)}} \phi_{m}^{(0)}
\eeq
Let us now try to make more precise calculations. The difference
from the textbook's solution comes due to the fact that we retain
the prescription (8). Then there is nothing to prevent one treating
the Hamiltonian (9) as the new {\bf unperturbed} Hamiltonian in
the new perturbation problem
\beq
(H_{1} + \lm^{2} W_{2} + \lm^{3} W_{3} + \lm^{4} W_{4} + O( \lm^5
) ) \phi^{(1)} = E \phi^{(1)}
\eeq
It is very important to notice that the order of the perturbation
in (12) is $ \lm^2 $ ( and higher powers ). The term $ O( \lm ) $
hidden in the Hamiltonian $ H_{1} $ is no more the small parameter.
This is in a sense "another $ \lm $ " as it is now a part of the
exactly solvable Hamiltonian $ H_{1} $. One may temporary call
it e.g. $ \rho $ in order not to confuse it with the real small
parameter $ \lm^{2} $. As we will see in a moment, this redefinition
leads to an infinite resummation of the Rayleigh - Schrodinger
perturbation series.

 We would like now to make again a canonical transformation such that
the transformed operators $ W_{2} , W_{3} $ etc. become diagonal
in the basis of the eigenstates of the Hamiltonian $ H_{1} $.
Obviously, this could be done with the choice  $ \phi^{(1)} =
exp( - \lm^2 S_{2} ) \phi^{(2)} $ ,
 then the term $ W_{2} $ is diagonalized.
However, the term $ W_{3} $ can be also diagonalized at the same
step. To this end, we have to choose
\beq
\phi^{(1)} = e^{ - \lm^{2} S_{2} - \lm^{3} S_{3}} \phi^{(2)}
\; , \; ( S_{2}^{+} = - S_{2} , S_{3}^{+} = - S_{3} )
\eeq
Let us show that with this choice we arrive at {\it independent}
equations for the operator $ S_{2} , S_{3} $ which are as simple as
eq. (7). ( an attempt to diagonalize also the term $ \lm^4 W_{4} $
would break this property ). We obtain the following equation for
the wave function $ \phi^{(2)} $ :
\bea
( H_{1} + \lm^{2} ( W_{2} + [ S_{2}, H_{1}] ) +
\lm^{3} ( W_{3} + [S_{3}, H_{1}] )  \nonumber \\
+ \lm^{4} ( W_{4} + [S_{2},W_{2}] + \frac{1}{2} [S_{2},[S_{2},
H_{1}]] )  + O( \lm^{5}) ) \phi^{(2)} = E \phi^{(2)}
\eea
Proceeding analogously to (7) ( and taking into account (10) ),
we obtain
\beq
\lno S_{2} \rmo = \frac{ \lno W_{2} \rmo }{ E_{n}^{(1)} - E_{m}^{(1)}}
\; , \;  \lno S_{3} \rmo = \frac{ \lno W_{3} \rmo }{ E_{n}^{(1)} -
E_{m}^{(1)}} \; , \; n \neq m
\eeq
(The tacit assumption we have done here is that the new perturbation
problem in not degenerate. If this is not the case, our procedure
terminates, and the degenerate perturbation theory must be used.
However, the degeneracy is not possible in the one-dimensional case,
provided the operators $ H_{1} , W_{i} $ are self-adjoint.)

  These canonical transformations can be continued. For example, at
the third step the terms up to $ \lm^{7} $ will be diagonalized in
respect to the eigenbasis of the Hamiltonian
\beq
H_{2} = H_{1} + \lm^{2} \tilde{W_{2}} + \lm^{3} \tilde{W_{3}}
\eeq
where $ \tilde{W_{2}} , \tilde{W_{3}} $ are the operator in
 front of the powers $ \lm^{2} , \lm^{3} $ in (14). What we
have obtained is just the quantum version of the superconvergent
procedure by Kolmogorov ( see Sect.1 and Sect.3 ). A few comments
are now in order.
\newline
(1) Transferring the averaged terms from the perturbation to
the new unperturbed Hamiltonian is the most non-trivial part
of our scheme. This is the precise analogue of the Newton method
and the Kolmogorov approach in classical mechanics. The only
difference is in the averaging procedure. We will see in the next
sections that this quantum averaging can be done formally identical
to the classical one.
\newline
(2) The suggested method corresponds to the infinite resummation
of the usual perturbative series. Let us illustrate this obvious
observation on the example of the $ O( \lm^{4} ) $ results. We
obtain from (13) , (14) :
\beq
\phi_{n}^{(1)} = \phi_{n}^{(0)} + \lm^{2} \sum_{ m \neq n}
\frac{ \lmo W_{2} + \lm W_{3} \rno }{ E_{n}^{(1)} - E_{m}^{(1)}}
+ \ldots
\eeq
We would like to emphasize once again that while we expand the
nominator to the given accuracy, retaining of all powers of $ \lm $
in the denominator is completely legitimate, as they are coming from
the exactly solvable problem with the Hamiltonian $ H_{1} $.
Let us also show that rational functions of the coupling constant
appear in the formulas for the energy levels starting from the third
step. The equation to be solved is
\beq
( H_{2} + \lm^{4} \tilde{W_{4}} + \ldots ) \phi^{(2)} =
E \phi^{(2)}
\eeq
Proceeding as previously, we substitute $ \phi^{(2)}  = \exp{ (- \lm^{
4} S_{4} + \ldots ) } \phi^{(3)} $ . Then the corrections to the energy
of order $ O( \lm^{4} ) $ are expressed in terms of the matrix elements
of the operator $ \tilde{W_{4}} $ ( see (14) ). Thus we immediately
observe an appearance of $ \lm $ in denominators owing to (15) .
In effect, we get automatically a Pade-type series for the
energy levels. It will be shown in Sect.5 that the corresponding
series is {\it convergent} for the quantum anharmonic oscillator.
\newline
(3) It can happen that at some step our equation for the operators
$ S_{i} $ will not possess anti-hermitean solutions, but instead will
give raise to some symmetric operators. Then a care must be taken
to define self-adjoint extensions of operators at hand since
otherwise the eigenvalues would not be preserved by a new canonical
transformation.
\newline
(4) The practical inconvenience of the method as it stands
is related to the necessity to re-calculate at every step the
matrix elements from one to another basis. This problem is quite
similar to an analogous problem in the classical perturbation theory
\cite{Lic} where the generating function is a function of both the old
and new variables while the new Hamiltonian must be expressed
as a function of the new variables only. An elegant solution to
this problem consists in proceeding to a "non-coordinate" description
of the canonical transformations which is called the Lie transformation
method ( see Sect.3 ). More important is the fact that in such a form,
the method admits a straightforward generalization to an arbitrary
dimensional case.
\newpage

\section{Superconvergence with Lie transforms}

The classical perturbation theory reduces the formal order of the
perturbation by virtue of successive canonical transformations
chosen such that to eliminate the phases (this procedure is to be
done with preventing secular terms, see \cite{Arn}, \cite{Lic}). A
generating function $ S( I_{1} , \theta , t) $ contains the
old ($ \theta $) and the new ($ I_{1} $) variables, thus a relation
between the new Hamiltonian $ \bH (I_{1}, \theta_{1} , t) $ and the
old Hamiltonian $ H ( I , \theta , t) $ appears also in a mixed
representation :
\beq
\bH (I_{1},\theta_{1} , t) = H ( I , \theta) + \frac{ \partial S(
I_{1}, \theta ,t)}{ \partial t}
\eeq
where
\beq
I = \frac{ \partial  S}{ \partial \theta} \;\; and \; \; \theta_{1} =
\frac{ \partial  S}{ \partial I_{1}}
\eeq
Eq. (19) is just the relation between the {\it values} of two
Hamiltonians at corresponding points in the phase space. In order
to convert it into a relation between the {\it functions}, one has
to solve the functional equations (20). This is not very convenient
when one goes beyond the leading order. An elegant modified version
has been proposed by Hori and Deprit (see \cite{Lic} , \cite{Hor} and
references
therein). In this version the transformation from the
old canonical variables $ \bx = (p_{i}, q_{i}) $ to the new canonical
$ \bbx = (\bar{p_{i}} , \bar{q_{i}} ) $  is given by a {\it Lie
generator} (rather than generating function) $ w( \bx ,\lm) = w_{1}
+ \lm w_{2} + \ldots $ such that the transformation  $ \bx \rightarrow
\bbx ( \bx ,\lm) $ is just the shift by the "time" $ \lm $ along the
trajectories of the system with the "Hamiltonian" $ w $ :
\beq
\frac{d \bbx}{d \lm} = [ \bbx , w ]
\eeq
(we use the same marking for the Poisson bracket in (21) as for
commutators in Sect.2 because, as will be clear soon, the formal
structure of the corresponding classical transformation turns out
to be identical to that of the quantum problem). One further
introduces the {\it Lie operator}  $ L $ , simply related to
the {\it vector field} $ X_{w} $ associated with the Hamiltonian
$ w $ :
\beq
L = [w , \; \; ] = \sum_{k} ( \frac{\partial w }{\partial q_{k}}
\frac{ \partial }{\partial p_{k}} - \frac{\partial w}{
\partial p_{k}} \frac{ \partial }{
\partial q_{k}} ) \equiv  - X_{w}
\eeq
and the {\it evolution operator} $ T $ which transforms any function
$ g $ at the new point $ \bbx ( \bx , \lm) $ into another function
$ f $ at the original point $ \bx $ :
\beq
f[ \bx] = g[ \bbx ( \bx ,\lm ) ] \; \; \Longleftrightarrow \; \; f = T g
\; \; ; \; \; \bx = T \bbx
\eeq
The operator $ T $ is exactly what is needed in order to convert (19)
into a functional equation since (23) means, in particular, that
formally
\beq
\bH = T^{-1} H
\eeq
(we are only interested here in time-independent problems. For
non-autonomous systems, the form (24) is incorrect, see \cite{Lic},
\cite{Hor}). The operator solution for $ T $ can be readily
deduced from the above formulas ; from (21) and (23) we
obtain
\beq
\frac{ d T} { d \lm} = - T L  \; \; \Longrightarrow \; \; T = \exp [
- \int^{ \lm}
L( \lm' ) d \lm' ]
\eeq
To see the formal similarity of the presented formulas with those
of quantum canonical transformation \cite{Eck}, let us retain for
simplicity only the leading term of the power expansion $ L = L_{1}
+ \lm L_{2} + \ldots $ in (24). Then (see (35) below)
\beq
\bbx = T^{-1} \bx = ( 1 + \lm L_{1} + \frac{1}{2} \lm^{2} L_{1}^{2}
+ \ldots ) =
e^{  \lm L_{1}} \bx
\eeq
The inverse transformation is
\beq
\bx = T \bbx = ( 1 - \lm L_{1} + \frac{1}{2} \lm^{2} L_{1}^{2} -
 \ldots ) \bbx = e^{ - \lm L_{1}}  \bbx
\eeq
The Hamiltonian equations of motion written via the Hamiltonian
vector field $ X_{H} $
\beq
\frac{ d \bx }{d t} = X_{H} \bx \equiv - \sum_{k} ( \frac{ \partial H}{
\partial
q_{k}} \frac{ \partial }{ \partial p_{k}} - \frac{\partial H}{
\partial p_{k}} \frac{
 \partial }{ \partial q_{k}} ) \bx
\eeq
now read (we suppose L to be independent of time)
\beq
e^{ - \lm L_{1}} \frac{ d \bbx}{ d t} = X_{H} e^{ - \lm L_{1}} \bbx
\eeq
or
\beq
\frac{ d \bbx}{ d t} = X_{ \bH} \bbx
\eeq
where
\beq
X_{ \bH} = e^{ \lm L_{1}} X_{H} e^{ - \lm L_{1}} =
 \sum_{n=0}^{ \infty} \frac{ \lm^{n}}{ n!} \underbrace{
[ L_{1},[L_{1}, \ldots [ L_{1}, X_{H}]]] }_{n \; times}
\eeq
(the brackets mean the commutation operation when stand with vector
fields)
\newline
Using the identity $ [X_{w} ,X_{H}] = - X_{ [ w, H ] } $ \cite{Arn}
, we obtain from (31) the transformed Hamiltonian
\beq
\bH = \sum_{n=0}^{ \infty} \frac{ \lm^{n}}{ n!} \underbrace{ [w ,
 [w ,
\ldots [w , H ]]] }_{n \; times}
\eeq
The formal identity with the quantum mechanical formulas becomes
quite transparent. All formulas of this section will be true in
the quantum case, provided one substitutes "Poisson brackets"
$ \rightarrow $ "quantum commutators" and "functions on the phase
space" $ \rightarrow $ "operators in the Hilbert space".
\newline
To get the perturbative series, one further expands
\bea
w &=& \sum_{n=0}^{\infty} \lm^{n} w_{n+1} \; \; \; ; \; \; \;
L = \sum_{n=0}^{\infty} \lm^{n} L_{n+1} \nonumber \\
T &=& \sum_{n=0}^{\infty} \lm^{n} T_{n} \; \; \; ; \; \; \;
T^{-1} = \sum_{n=0}^{\infty} \lm^{n} T_{n}^{-1}  \\
H &=& \sum_{n=0}^{\infty} \lm^{n} H_{n} \; \; \; ; \; \; \;
\bH = \sum_{n=0}^{\infty} \lm^{n} \bH_{n} \; \; ,  \nonumber
\eea
where
\beq
L_{n} = [ w_{n} , \; \; ] \; \;  ; \; \; T_{0} = 1 \; \; ; \; \;
T_{0}^{-1} = 1
\eeq
Then a simple algebra leads to the recursion relation (see \cite{Lic} )
\beq
T_{n}^{-1} = \frac{1}{n} \sum_{m=0}^{n-1} L_{n-m} T_{m}^{-1}
\eeq
and, as the consequence of (24),  ( $ n > 0 $ )
\beq
D_{0} w_{n} = n ( \bH_{n} - H_{n} )  - \sum_{m=1}^{n-1} ( L_{n-m}
\bH_{m} + m T_{n-m}^{-1} H_{m} ) \; ,
\eeq
where
\beq
D_{0} = [ \; \; , H_{0} ] \; \; \; and \; \; \;  \bH_{0} = H_{0}
\eeq
To the fourth order, the above equation yields
\bea
D_{0} w_{1} &=& \bH_{1} - H_{1}  \\
D_{0} w_{2} &=& 2 ( \bH_{2} - H_{2}) - L_{1} ( \bH_{1} + H_{1}) \\
D_{0} w_{3} &=& 3 ( \bH_{3} - H_{3}) - L_{1}( \bH_{2} + 2 H_{2})
                - L_{2}( \bH_{1} + \frac{1}{2} H_{1}) - \frac{1}{2}
                L_{1}^{2} H_{1}        \\
D_{0} w_{4} &=& 4 ( \bH_{4} - H_{4}) - ( \frac{1}{3} L_{3} + \frac{1}{6}
               L_{1} L_{2} + \frac{1}{3} L_{2} L_{1} + \frac{1}{6}
               L_{1}^{3} ) H_{1}  \\
            &-& L_{3} \bH_{1} - ( L_{2} + L_{1}^{2}) H_{2}
                - L_{1} \bH_{3} - 3 L_{1} H_{3} - L_{2} \bH_{2}
\eea
The standard (von Zeipel's)
perturbation theory is reproduced in this way as follows.
Given $ H_{1} $, one chooses $ \bH_{1} $ such that secularities in the
r.h.s. of (38) are eliminated, and then finds $ w_{1} $. At second order,
one substitutes the found $ w_{1} $ into (39) and looks for a $ w_{2} $
which eliminates secularities in the r.h.s. of (39), etc.
\newline
As has been stressed in \cite{Lic}, the Lie transforms formalism
is more convenient for elucidating the superconvergence than the
standard method of canonical transformations. The only difference
from the above described line of reasoning is a different choice
of the Lie generators $ w_{i} $.
\newline
The principal rule is exactly parallel to the method used by us
in Sect.2 : successive new Hamiltonians are obtained by successive
Lie transforms while equations for different $ w_{n}^{(i)} $ and
$ w_{m}^{(i)} $ ( the superscript $ (i) $ will generally denote
the number of the iteration ) must be independent of each other.
At each step, chosen as many as possible independent $ w^{(i)} $'k,
all remaining $ w^{(i)} $'s are set equal to zero. We describe two
steps of the Kolmogorov scheme following \cite{Lic}.
\newline
At the first step, there is only one equation (38). To kill the
secular term, we set $ \bH_{1}^{1} = \;
< H_{1} > $ ($ < \;> $ here denotes
averaging over the "fast" angle variables. In Sect. 4 we will,
however, follow another averaging procedure ). Then $ w_{1}^{(1)} $
is determined from
\beq
D_{0} w_{1}^{(1)} = - \{ H_{1} \}  \; \; \; ; \; \; \;
                     w_{i}^{(1)} = 0 \; , \; i > 1
\eeq
where $ \{ \; \; \} $ denotes the oscillating part. All others
$ \bH_{i}^{(1)} $ are given by (39 - 42) with the constraint (43).
One finds
\bea
\bHo_{2} &=& \frac{1}{2} [ w_{1}^{(1)} , ( \bHo_{1} + H_{1})] \\
\bHo_{3} &=& \frac{1}{3} [ w_{1}^{(1)} , ( \bHo_{2} + \frac{1}{2}
    [ w_{1}^{(1)} , H_{1}] ) ]
\eea
One can show that the general recursion formula reads ( $ n > 2 $ )
\beq
\bHo_{n+1} = \frac{2}{n+1} [ \wo_{1} , ( \bHo_{n} - \frac{1}{2 n}
 [ \wo_{1} , \bHo_{n-1} ] ) ]
\eeq
At the second step, we absorb the averaged (i.e. solvable) part
$ \lm <H_{1} > $ into the zero order part of the new "old
Hamiltonian" $ \Ho $ :
\bea
\Ho_{0} &=& \bHo_{0} + \lm <H_{1} >  \nonumber \\
\Ho_{1} &=& 0            \\
\Ho_{i}  &=& \bHo_{i} \; \; \; , \; \; \; i > 1   \nonumber
\eea
After the second transformation $ \wt $, we obtain a new Hamiltonian
$ \bHt $ with $ \bHt_{0} = \Ho_{0} $. As there is no first-order
perturbation, we may choose $ \wt_{1} = 0 $. Thus, we obtain from
(39 -42) two independent equations for $ \wt_{2} $ and $ \wt_{3} $ :
\bea
D_{0}^{(1)} \wt_{2} &=& 2 ( \bHt_{2} - \Ho_{2} )  \\
D_{0}^{(1)} \wt_{3} &=& 3 ( \bHt_{3} - \Ho_{3} )
\eea
while
\beq
D_{0}^{(1)} \wt_{4} = 4 ( \bHt_{4} - \Ho_{4} )
   - [ \wt_{2} , ( \bHt_{2} + \Ho_{2} ) ]
\eeq
where the new time derivative is calculated in respect to the
zero order new "old" Hamiltonian :
\beq
D_{0}^{(1)} = [ \; \; \; , \Ho_{0} ]
\eeq
The equation (48),(49) are solved simultaneously choosing $ \bHt_{2} $
and $ \bHt_{3} $ to eliminate secularities in $ \wt_{2} $ and $ \wt_{3} $
, as has been done in (43). All $ \wt_{i} \;  , \; i > 3 $ are set
equal to zero. Then the next terms in the transformed Hamiltonian  are
\bea
\bHt_{4} &=& \Ho_{4} + \frac{1}{4} [ \wt_{2} , \bHt_{2} +
          \Ho_{2} ]           \\
\bHt_{n} &=& \Ho_{n} + \frac{1}{n} \sum_{m=1}^{n-1} ( L_{n-m}^{(2)}
 \bHt_{m} + m T_{n-m}^{(2) \; -1} \Ho_{m} )
\eea
At the third step, the new "old Hamiltonian" is built analogously
to (47) :
\bea
\Ht_{0} &=& \bHt_{0} + \lm^{2} < \Ho_{2} > + \lm^3 < \Ho_{3} >
  \nonumber  \\
\Ht_{i} &=& 0  \;  \; , \; \; i = 2 \; , \; 3   \\
\Ht_{i} &=& \bHt_{i} \; \; , \; \; i > 3     \nonumber
\eea
This iterative procedure can be continued. It can be shown by
induction \cite{Lic} that at the nth step, the terms from
$ i = 2^{n} $ to $ i = 2^{n+1} - 1 $ are diagonalized. This
completes the description of the Kolmogorov method within
the Lie transforms formalism.
\newline
It should be mentioned that, while being extremely important
theoretically (for the proof of the KAM theorem), the Kolmogorov
method did not get much practical applications. The only known to
the author example has been given by Chirikov \cite{Chir} who has
shown how to make the first Kolmogorov's step in the problem
of the classical one-dimensional pendulum using the usual formalism
of canonical transformations. But, as we have seen in Sect.2 ,
really interesting phenomena  within the Kolmogorov method start
only from the second step. It is also worth noting that the
notion of superconvergence has been used in a different content
in \cite{Eck} in treating the Birkhoff-Gustavson normal form.

\section{Classical anharmonic oscillator}

For this one-dimensional classical problem, the use of
the perturbation theory is nothing but a formal exercise. Indeed,
the corresponding equation of motion can be solved in terms of
 elliptic functions that indicates that a perturbative series
is convergent. Obviously, this convergence is related to the absence
of resonances in any one-dimensional system. The perturbative series
in this case is a formal expansion of the Hamiltonian in powers of
the harmonic oscillator Hamiltonian. A final answer is usually
expressed in the form of an expansion of the
frequency of non-linear oscillations in powers of the energy.
 We will study this problem  merely for the sake of comparison
and working out reliable tools for treating the corresponding
quantum problem.
\newline
We start with the Hamiltonian
\beq
H ( p ,x ) = \frac{1}{2} p^{2} + \frac{1}{2} \omega^{2} x^{2}
  + \lm \omega^{2} x^{4}
\eeq
It is very convenient to proceed from the canonical variables
$ ( p ,x ) $ to the new canonical complex-conjugate variables
\bea
z   &=& \frac{1}{ \sqrt{2 \omega}} ( \om x + i p )  \nonumber \\
\bz &=&  \frac{1}{ \sqrt{2 \om}} ( \om x - i p )
\eea
Then
\bea
H      &=&  H_{0} + \lm H_{1}  \nonumber \\
H_{0}  &=&  \om \bz z    \\
H_{1}  &=& \frac{1}{4} ( z + \bz )^{4} = \frac{1}{4} ( \bz^{4}
        +4 \bz^{3} z + 6 \bz^{2} z^{2} + 4 \bz z^{3} + z^{4} )
\nonumber
\eea
In the complex variables the Poisson bracket takes the form
\beq
[ f , g ] = - i \{  \frac{ \partial f }{ \partial z } \frac{
\partial g }{ \partial \bz}  - \frac{ \partial f}{ \partial \bz}
\frac{ \partial g }{ \partial z }  \}
\eeq
The advantage of proceeding to ($ z , \bz $) and not to the
action-angles variables is three-fold : (1) it makes quantum
mechanical formulas written in the second quantized form
very similar to classical ones;
(2) simple explicit formulas for the Lie generators to all
orders can be given
                         ; and (3) the averaging procedure
       in the complex variables becomes much simpler in
comparison with averaging in the action-angle variables.
Instead of integration over the angles, the averaging is simply
given by picking up terms with equal powers of $ \bz $ and $ z $
in a product. This procedure is obviously equivalent to the
integration over the angle variables.
\newline
Proceeding along the line of Sect.3 , we obtain at the first
step
\bea
\bHo_{0} &=& H_{0} = \om \bz z  \nonumber \\
\bHo_{1} &=& < H_{1} > = \frac{3}{2} \bz^{2} z^{2}
\eea
Solving eq. (43) yields
\beq
\wo_{1} = i \frac{1}{ 4 \om} ( \frac{1}{4} \bz^{4} + 2 \bz^{3} z
- 2 \bz z^{3}  - \frac{1}{4} z^{4}  )
\eeq
The transformed Hamiltonian is calculated according to (44),(45)
(46) :
\bea
\bHo_{2} &=& \frac{1}{8 \om } [ - 34 \bz^{3} z^{3} + (
   \bz^{6}  - 6 \bz^{5} z - 33 \bz^{4} z^{2}  + ( \; c. \; c. \;)
    )  ]    \\
\bHo_{3} &=& \frac{1}{ 96 \om^2} [ 2250 \bz^{4} z^{4} +
( 33 \bz^{8} + 72 \bz^{7} z  + 348 \bz^{6} z^{2} + 1272 \bz^{5}
z^{3}  + ( \;c. \; c. \;) \; ) ]      \\
\bHo_{4} &=& \frac{3}{256 \om^{3}} [ -5244 \bz^{5} z^{5} +
( 37 \bz^{10} - 10 \bz^{9} z - 435 \bz^{8} z^{2} - 1760 \bz^{7} z^{3}
 \nonumber  \\
         &-& 4210 \bz^{6} z^{4}  + ( \; c. \; c. \; ) \; ) ]
\eea
where $ ( \; c. \; c. \;) $ means the complex conjugation.
The lengthy intermediate calculations have been made with the help
of the {\it Mathematica} package of analytic calculations.
Proceeding to the second step, the new "old" Hamiltonian becomes
\bea
\Ho_{0} &=& \om \bz z + \lm \frac{3}{2} \bz^{2} z^{2} \nonumber \\
\Ho_{1} &=& 0  \\
\Ho_{i} &=& \bHo_{i} \; \; , \; \; i > 1  \nonumber
\eea
We choose
\bea
\bHt_{2} &=& < \Ho_{2} > = - \frac{17}{4 \om} \bz^{3} z^{3}  \\
\bHt_{3} &=& < \Ho_{3} > = \frac{375}{16 \om} \bz^{4} z^{4}
\eea
The equations (48),(49) are of the form
\beq
[ \wt_{n} , \Ho_{0} ] =  \om [ \wt_{i}, \bz z + \frac{3 \lm}{ 2 \om}
\bz^{2} z^{2} ] = -  n \{ \Ho_{n}  \}
\eeq
To solve these functional equations, we use the following formula
\beq
[ \frac{ \bz^{m} z^{n}}{ 1 + \frac{3 \lm}{ \om} \bz z} \; ,\; \Ho_{0} ]
 = i \om ( m - n)  \bz^{m} z^{n}
\eeq
which can by checked by the direct calculation. Then answers for
$ \wt_{2} \; , \wt_{3} $
read
\bea
\wt_{2} &=& i \frac{1}{ 4 \om^{2}} \frac{1}{ 1 +
\frac{3 \lm}{ \om} \bz z} [ \frac{1}{6} \bz^{6} - \frac{3}{2}
\bz^{5} z  - \frac{33}{2} \bz^{4} z^{2} - ( \: c. \; c. \; ) ] \\
\wt_{3} &=& i \frac{1}{ 48 \om^{3}} \frac{1}{ 1 + \frac{3
\lm}{  \om} \bz z } [ \frac{33}{8} \bz^{8} + 12 \bz^{7} z +
87 \bz^{6} z^{2} + 636 \bz^{5} z^{3} - ( \; c. \; c. \; ) ]
\eea
We give the final form of the Hamiltonian to the $ O( \lm^{7} ) $
accuracy ( Here $ I $ stands for the product $ \bz z $ )  :
\bea
H^{(3)} &=& \om I + \frac{3}{2} \lm I^{2}
- \frac{17}{4} \frac{ \lm^{2}}{
\om} I^{3} + \frac{375}{16} \frac{ \lm^{3}}{ \om^{2}} I^{4}  \nonumber \\
   &-& \frac{15}{320} \frac{ \lm^{4}}{ \om^{3}} I^{5} \frac{ 3563 +
13496 \frac{ \lm}{ \om} I + 11799 \frac{ \lm^{2}}{ \om^{2}} I^{2} }{ (1
+ 3 \frac{ \lm}{ \om} I )^{2} }            \\
   &+& \frac{1}{320} \frac{ \lm^{5}}{ \om^{4}} I^{6} \frac{ 285299
  + 949266 \frac{ \lm}{ \om} I + 566055 \frac{ \lm^{2}}{ \om^{2}}
 I^{2} }{ ( 1+ 3 \frac{ \lm}{ \om} I )^{2} }  \nonumber  \\
   &-& \frac{1}{512} \frac{ \lm^{6}}{ \om^{5}} I^{7}  \frac{ 4239854
+ 33800451 \frac{ \lm}{ \om} I + 93912282 \frac{ \lm^{2}}{ \om^{2}}
I^{2} + 99162117 \frac{ \lm^{3}}{ \om^{3}} I^{3} +
22951350 \frac{ \lm^{4}}{ \om^{4}} I^{4}}{ (1
 + 3 \frac{ \lm}{ \om} I)^{4} }  \nonumber
\eea
Obviously, at the next steps we will get the successive zero order
Hamiltonians in the form of rational functions of $ \frac{ \lm}{ \om}
I $ . The corresponding $ w^{(i)} $'s can be calculated using the
following generalization of (68) :
\beq
[ \frac{ \bz^{m} z^{n}}{ P(\bz z)} \; , \; Q( \bz z) ]
= - i (n - m)  \bz^{m} z^{n}  \frac{ Q' ( \bz z)}{ P ( \bz z)}
\eeq
where $ P(y) \; , \; Q(y) $ are arbitrary functions, $ P(y) \neq 0 $
and the derivative $ Q'(y) $ exists. To solve the equation
\beq
[ w ( \bz , z ) \; , \; Q ( \bz z ) ] =  \frac{ \bz^{l} z^{k}}{
P_{1} ( \bz z)}
\eeq
appearing in higher orders, we have to choose in (72)
\beq
P ( \bz z) = Q'( \bz z) P_{1} ( \bz z)
\eeq
The equations (72-74) in principle solve the problem of
finding the corresponding Lie generators $ w^{(i)} $ at
any order. There are a few lessons to be learnt from the
above calculation. In the second quantization representation,
which is the relevant arena for building a corresponding
quantum scheme, one can expect somewhat analogous to (71)
expression for an approximate quantum Hamiltonian, the
action $ I $ being substituted by the particle number
operator $ N = a^{+} a $ , i.e. a kind of the operator
Pade expansion. The obtained Hamiltonian has to be checked
for being self-adjoint. Furthermore, to proceed beyond the
first step, quantum substitutes of eq. (68),(72) are needed.

\section{Quantum anharmonic oscillator}

The quantum anharmonic oscillator is the standard test problem
for any new quantum field theory - oriented method. A main
feature, that the anharmonic oscillator shares with field -
theoretical problems, is the zero radius of convergence of the
Rayleigh-Schrodinger perturbation series \cite{Ben} which
prevents an extension of perturbative results into the strong
coupling region. There has been a number of attempts to build
strong coupling expansions meant to be generalizable to the
field theory, refs. \cite{Hio} may probably be mentioned
among the most interesting ones.
\newline
Intuitively, the divergence of the perturbative series can be
understood as a result of the inability to reproduce a behavior
in the vicinity of the essential singularity at the origin
( the Dyson singularity ) in terms of an expansion in powers of
the coupling constant. However, it is well known that the
convergence can be
gained after the perturbative series is resumed,
the most popular resummation methods are due to Pade and Borel,
see e.g. \cite{Sim} for a review. In this section we would like
to argue that the quantum superconvergence leads to the
{\it uniform convergence} of the corresponding series for both
small and strong couplings due to the fact that our results
have the form of the (operator and functional) Pade series.
\newline
For the quantum version of the Hamiltonian (55) (with $ \om = 1 $)
we introduce the creation and annihilation operators :
\beq
a = \frac{1}{ \sqrt{2 \hbar}} ( \hat{x} + i \hat{p} ) \; \;
, \; \;
\bc = \frac{1}{ \sqrt{ 2 \hbar}} ( \hat{x} - i \hat{p} )
\eeq
and the particle number operator $ N = \bc a $ satisfying the
commutation relations
\beq
 [ a \; , \; \bc^{n} ]   = n \bc^{n-1}
 \; \; , \; \;  [ a^{n} \; , \; \bc ] =
  n a^{n-1}
\eeq
The unperturbed and perturbating Hamiltonians are
\bea
H_{0} &=& \hbar  ( \bc a + \frac{1}{2} )  \\
H_{1} &=& \frac{ \hbar^{2}}{ 4 } [ 3 ( 1 + 4 \bc a +
2 \bc^{2} a^{2} ) + ( \bc^{4} + 4 \bc^{3} a + 6 \bc^{2} + (h.c.) ) ]
                     \nonumber
\eea
where $ (h.c.) $ means the hermite conjugation. Proceeding analogously
to the classical case, we obtain $ \bHo_{1} = < H_{1} > $ and then
\beq
\wo_{1} = \frac{ \hbar}{ 4} ( \frac{1}{4} \bc^{4} + 2 \bc^{3} a
+ 3 \bc^{2} - (h.c.) )
\eeq
The transformed Hamiltonian , according to (44-46), is
\bea
\bHo_{2} &=& \frac{ \hbar^{3}}{16} [ -42 - 288 \bc a - 306 \bc^{2} a^{2}
-68 \bc^{3} a^{3} +  \ldots ]  \nonumber \\
\bHo_{3} &=& \frac{ \hbar^{4}}{16} [ 333 + 3582 \bc a + 6291 \bc^{2}
a^{2} + 3000 \bc^{3} a^{3} + 375 \bc^{4} a^{4} + \ldots ]  \\
\bHo_{4} &=& \frac{ \hbar^{5}}{256} [ - 26649 - 390366 \bc a -
977562 \bc^{2} a^{2} - 741636 \bc^{3} a^{3} - 196650 \bc^{4} a^{4}
- 15732 \bc^{5} a^{5} + \ldots ]     \nonumber
\eea
where we have explicitly shown only the diagonal parts of the
operators, the complete forms are rather
lengthy.             Proceeding to the second step, we add the
diagonal part of the perturbation $ H_{1} $ to the unperturbed
Hamiltonian $ H_{0} $ to get the new "unperturbed Hamiltonian",
just as we have done in the classical treatment :
\bea
\Ho_{0} &=& \hbar ( N + \frac{1}{2} ) + \frac{ 3 \lm
\hbar^{2}}{4} ( 1 + 2 N  + 2 N^{2})   \nonumber \\
\Ho_{1} &=& 0            \\
\Ho_{i} &=& \bHo_{i} \; \; , \; \; i > 1    \nonumber
\eea
At the second step, the quantum analogue of eq. (67) can be written
as
\beq
[ N + k N^{2} \; , \; \wt_{2} ] = \frac{ 2 ( 1 -k )}{ \hbar} \{ \Ho_{2}
\}
\eeq
where we have denoted
\beq
k = \frac{ \frac{3}{2} \lm \hbar}{ 1 + \frac{3}{2} \lm \hbar}
\eeq
The quantum version of eq. (68) reads
\beq
[ N + k N^{2} \; , \; \frac{1}{ 1 - k (n - m) + 2 k N } \bc^{n} a^{m} ]
= ( n - m) \bc^{n} a^{m}
\eeq
The above formula yields the following answer for the Lie generator
$ \wt_{2} $
\bea
\wt_{2} &=& \frac{(1- k) \hbar^{2}}{8} [ \frac{1}{3} \frac{1}{ 1
- 6 k + 2 k N } \bc^{6} - \frac{15}{2} \frac{1}{ 1 - 4 k + 2 k N}
\bc^{4} - \frac{171}{2} \frac{1}{ 1 - 2 k + 2 k N} \bc^{2} \nonumber \\
        &-&  \frac{132}{ 1 - 2 k + 2 k N} \bc^{3} a -
    - \frac{3}{ 1 - 4 k + 2 k N} \bc^{5} a  -
 \frac{33}{ 1 - 2 k + 2 k N} \bc^{4} a^{2} - (h.c.)  ]
\eea
The result for the Lie generator $ \wt_{3} $ has a similar structure.
To calculate higher order terms of the transformed Hamiltonian, one
needs a few more commutators which can be obtained as the simple
consequences of the Haussdorff identity and Schwinger exponential
representation of operators ( $ \alpha $ and $ \beta $ are
arbitrary c-numbers ) :
\beq
[ \bc^{n} a^{m} , \frac{1}{ \alpha + \beta N} ] =  ( \frac{1}{ \alpha
- \beta ( n - m ) + \beta N} - \frac{1}{ \alpha + \beta N } )
\bc^{n} a ^{m}
\eeq
\beq
[ a^{m} , \bc^{n} ] = n! \; \bc^{n - m} \sum_{l = max \{ 0 , m -n \}
}^{m -1} \frac{ m!}{ l! (m-l)! (n-m+l)! } \bc^{l} a^{l}
\eeq
Using these commutators, one can easily check that the operator
$ \wt_{2} $ is indeed anti-hermitean.
As the resulting expression for the second-iteration Hamiltonian
is rather lengthy, we will retain below only the terms possessing
non-zero vacuum expectation values :
\bea
H^{(3)} &=& \frac{ \hbar}{2} + \frac{3}{4} \lm \hbar^{2}
- \frac{21}{8} \lm^{2} \hbar^{3} + \frac{333}{16} \lm^{3} \hbar^{4}
- \frac{26649}{ 256} \lm^{4} \hbar^{5}  \nonumber   \\
&-& \frac{ \lm^{4} \hbar^{5} ( 1 -k)}{16} [
\frac{29241}{16}
\frac{1}{1 + 2 k + 2 k N} + \frac{675}{2} \frac{1}{1 + 4 k + 2 k N}
+ \frac{30}{1 + 6 k + 2 k N} ] \nonumber \\
&+& \frac{444609}{512} \lm^{5} \hbar^{6} + \frac{ \lm^{5} \hbar^{6}
(1 - k)}{4} [ \frac{208791}{32} \frac{1}{ 1 + 2 k + 2 k N} \\
&+& \frac{14985}{8} \frac{1}{1+ 4 k + 2 k N} - \frac{315}{1+
6 k + 2 k N} ]   + \ldots   \nonumber
\eea
One can easily see that the perturbative expansion
for the ground state energy \cite{Ben}
\beq
E_{0}= \frac{ \hbar}{2} + \frac{3}{4} \lm \hbar^{2}
- \frac{21}{8} \lm^{2} \hbar^{3} + \frac{333}{16} \lm^{3} \hbar^{4}
- \frac{30885}{128} \lm^{4} \hbar^{5} + \frac{916731}{256} \lm^{5}
\hbar^{6} + \ldots
\eeq
is reproduced from our formula by re-expanding the rational fractions.
For example, about 25 \% of the perturbative result at the $ \lm^{5} $
order is due to the expansion of our $ O( \lm^{4} ) $ fraction.
It seems plausible to suggest that the Kolmogorov series reproduces
the usual perturbative result at all orders, though we do not get
a general proof.
\newline
Heuristically, one can expect the convergence of the Kolmogorov
series (87) for the ground state energy due to the fact that, the higher
is the order of the perturbative coefficient, the larger number of
fractions will contribute to its value.
This statement can be given a formal proof. The resulting from
our procedure fractions can be reorganized into rational functions
of the coupling constant. A convenient choice is e.g. to collect
into a single rational function all fractions resulting from one
Kolmogorov's step. Let us suppose that at some step we have obtained
the answer in the form
\beq
P_{ N_{1} , j_{1}}  ( \lm ) \equiv  f^{ [ N_{1} , N_{1} + j_{1} ] }
( \lm ) =  R^{ [ N_{1} + j_{1} ] } / Q^{ [ N_{1} ] }
\eeq
where $ R^{ [N] } ( Q^{ [M] } )  $ is a polynomial of degree N (M).
(note that in our formulas $ j_{1} > 0 $) and at the next step
the answer is of the form
\beq
P_{ N_{2} , j_{2}}  ( \lm ) \equiv  f^{ [ N_{2} , N_{2} + j_{2} ] }
( \lm ) \; \; , \; \; ( N_{2} > N_{1} )
\eeq
Let us prove that $ j_{1} = j_{2} $. Suppose this were not true.
As $ P_{N_{1},j_{1}} \; , \; P_{N_{2},j_{2}} $ can be thought
as the representatives of two series of the Pade approximants,
there exist the limits \cite{Sim}
\beq
f_{ j_{1}}  ( \lm ) = \lim_{ N \rightarrow \infty}
f^{ [ N , N + j_{1} ] }  \; \; \; , \; \; \;
f_{ j_{2}}  ( \lm ) = \lim_{ N \rightarrow \infty}
f^{ [ N , N + j_{2} ] }
\eeq
and the convergence is uniform in the cut $ \lm $ plane. However,
$ f_{j_{1}} = f_{j_{2}} $ since a measure giving the perturbative
coefficients
in the form of its moments is unique for the anharmonic oscillator
problem \cite{Sim}. Then it means that the series $ f^{ [ N ,
N + j_{1} ]} - f^{ [ N , N + j_{2} ]} $ converges uniformly to 0.
As $ j_{1} \; , \; j_{2} > 0 $ , this is only possible when
$ j_{1} = j_{2} $ (at very large $ \lm  \; \;  f^{ [ N , N +
j] } \sim  \lm^{j} $ ). The fact that
$ j_{1} = j_{2} $  means that the Pade series obtained from
Kolmogorov procedure is diagonal and therefore converges
uniformly to the true eigenvalue \cite{Gra}. (Note that we have
explicitly checked that all poles resulting from the Kolmogorov series
are located at real negative $ \lm $ for any energy level and
the resulting Hamiltonian is self-adjoint.)
\newline
We would like to end up this section with a few comments.
 At any order, the eigenfunctions of a resulting effective
Hamiltonian are the harmonic oscillator wave functions in proper
coordinates. The proposed approach is essentially the method
of a free field representation. Given the needed accuracy, the
interaction is transferred to such canonical transformation
that yields the free (i.e. harmonic oscillator)  Hamiltonian.
The perturbed wave function in the initial coordinates is given
by the inverse transformation of the basis. The theory of
linear canonical transformations in the Bargmann-Fock space
is well studied \cite{Bargm}. As we deal with non-linear
canonical transformations, more sophisticated methods are needed.
We feel that a method suggested recently \cite{And} can be of use here.
Note also that the corresponding treatment in a many-dimensional case
provides a non-linear extension of the famous Bogolyubov transformations.
\newline
The results obtained so far can be summarized as follows.
In all cases where the usual perturbation theory is applicable,
the superconvergent methods corresponds to a Pade-type
resummation of the perturbative series. In the above studied
case, the convergence of the Kolmogorov expansion is the
consequence of general theorems on the Pade approximation \cite{Sim}.
In other problems, convergence properties have to be tested
separately. Instead of discussing this topic, in the next section
we will show how a similar re-organization of the perturbation theory
allows one to treat approximately a more complicated system with
two degrees of freedom, where the
usual perturbation theory cannot be applied.

\section{Yang-Mills quantum mechanics}

Yang-Mills (YM) classical mechanics has been proposed \cite{Cha}
as a toy model for a study of non-perturbative nonlinear YM waves
in the Minkowsky space. There one starts with fields in the SU(2)
YM theory which depend only on the time (this is the reasonable
assumption in the infra-red). Then, after the gauge fixing
$ A_{0}^{a} = 0 $ , the classical YM theory turns out to be
essentially equivalent to the dynamical system with the Hamiltonian
\cite{Cha}
\beq
H = \frac{1}{2} \sum_{a= 1}^{3}
( \frac{ d f^{a}}{d t} )^{2} + \frac{1}{2}
\sum_{a=1}^{3} \sum_{b \neq a} (f^{a})^{2} (f^{b})^{2}
\eeq
where the dynamical variables $ f^{a} $ parametrize the gauge
fields $ A_{i}^{a} $ ( $ O_{i}^{a} $ is a time-independent
orthogonal matrix ):
\beq
A_{i}^{a} = O_{i}^{a} f^{a} (t) /g \; \; ( no \; \; sum \; \;
over \; \; a ) \; \; \; and \; \; \; O_{i}^{a} O_{i}^{b} =
\delta^{ab}
\eeq
A simplified system obtained from (92) by fixing $ f^{3} = 0 $ is
still highly non-trivial and commonly called YM classical mechanics:
\beq
H_{YM} = \frac{1}{2} ( p_{1}^{2} + p_{2}^{2} ) + \lm x_{1}^{2} x_{2}^{2}
\eeq
(we have introduced here the coupling constant $ \lm $ instead of
1/2 in (92) ).
\newline
In spite of its seeming simplicity, the Hamiltonian (94) leads to
a very complicated chaotic motion \cite{Cha}. It has been believed
for a long time that the system (94) is ergodic \cite{Cha}. More recently,
small stability islands have been found \cite{Dah}.
There also has been a considerable interest in a quantum system with
the Hamiltonian (94) \cite{Sav}, \cite{Bar}, \cite{SymYM}, \cite{Med},
\cite{Pol} (note that (94) is not the homogeneous-space model
of the quantum YM theory, but still is closely related to the latter
\cite{Sav}). The fact that a spectrum is discrete has been proved in
\cite{SymYM}. Different methods, based on the quasiclassical and
quantum adiabatic approximations \cite{Med}
 and summation of adiabatically stable periodic
orbits \cite{Pol}, have been proposed for finding the spectrum and
wave functions.
\newline
It is obvious beforehand that the usual perturbation theory is
hopelessly inapplicable for treating the system (94). No matter
how $ \lm $ is small, the perturbation results in a reconstruction
of the spectrum from continuous to the discrete one, the huge effect.
On contrary, a modified perturbation scheme a la Kolmogorov can be
proposed which allows one to build a controllable approximation to
the WF near the origin, where the adiabaticity breaks down (see below).
\newline
Let us consider the "perturbation" in the second quantization
representation (75) (we set $ \hbar = 1 $ and use $ z_{i} \; , \bz_{i} $
to denote the quantum annihilation and creation operators) :
\bea
\lm x_{1}^{2} x_{2}^{2} &=& \lm \frac{1}{ 4 } [ 1 + 2 N_{1}
+ 2 N_{2} + 4 N_{1} N_{2} ] + \lm \frac{ 1}{4} \{ z_{1}^{2}
z_{2}^{2} + z_{1}^{2} \bz_{2}^{2}  \nonumber \\
                        &+& ( 1 + 2 N_{2} ) z_{1}^{2} +
( 1 + 2 N_{1}) z_{2}^{2} + (h.c.) \}
\eea
In obtaining (95), we have used the
non-commutativity of the quantum operators, that has allowed one
to extract from the "perturbation" the part which is actually not
a perturbation as it is diagonal in the oscillator basis.
The expression denoted by [...] in (95) must be treated as a part
of an unperturbed Hamiltonian, if we intend to end up with a discrete
spectrum. The fact that the "unperturbed Hamiltonian" is of the same
order $ \lm $ as the remaining "perturbation" is not an obstacle
since after the first step the perturbation is only of order
$ \lm^{2} $(plus a small part of order $ \lm $ ). An additional problem
arises due to the fact that
both the unperturbed Hamiltonian and perturbation are degenerate so that
the degenerate perturbation theory also cannot be used. We propose
the following simple trick to handle this problem : let us just make
both the unperturbed Hamiltonian and perturbation non-degenerate.
We choose to transfer into the unperturbed Hamiltonian the
terms $ 1 + 2 N_{1} + 4 N_{1} N_{2} $ while the term $ 2 N_{2} $ is
retained as the part of the perturbation. The concrete choice $
N_{1} \leftrightarrow N_{2} $ does not matter as the difference may
be thought as the one to be repaired by higher order corrections.
In other words, we sacrifice the exact symmetry of the problem
in favor of only approximate one. (It is known that such procedure can yield
wrong results for the large distance behavior, but, as will be shown below,
in this region a different approximation becomes operative).
 As will be seen in a moment, this
suggestion indeed leads to avoiding small denominators at the lowest
order. Moreover, in such form the Hamiltonian is suitable for working
out the Kolmogorov scheme, as the new perturbation contains the part
which has a non-zero average (the $ 2 N_{2} $ term).
\newline
Adding to (95) the kinetic term operators, we arrive at the Hamiltonian
\bea
H     &=& H_{0} + \var H_{1}      \nonumber \\
H_{0} &=& \omo N_{1} + \omt N_{2} + \mu N_{1} N_{2} + \alpha   \\
H_{1} &=& \sum_{n,m = 0} ( 2 N_{2}  +   \bzo^{n} \bzt^{m} f_{nm}+
+ f_{nm}^{ \ast} \zo^{n} \zt^{m} +
\bzo^{n} g_{nm} \zt^{m} + \zo^{n} g_{nm}^{ \ast} \bzt^{m} )
\nonumber
\eea
where
\bea
\omo    &=& \frac{1}{2} + \frac{ \lm}{2} \; \;
  ,    \; \; \omt  = \frac{1}{2}  \; \;
, \; \;    \mu  = \lm     \nonumber \\
\alpha  &=& \frac{1}{2} + \frac{ \lm}{4} \; \;
, \; \;    \var= \frac{ \lm}{4}   \; \;
, \; \; \gamma     =  1 - \frac{1}{ \lm }
\eea
and the only non-zero $ f 'k \; , \;  g 'k $  are
\beq
f_{02} = \gamma + 2 N_{1} \; \; , \; \; f_{20} = \gamma + 2 N_{2} \; \; ,
\; \; f_{22} = 1 \; \; , \; \; g_{22} = 1
\eeq
(Note that the Hamiltonian of so-called Yang-Mills-Higgs quantum
mechanics \cite{Sav}, which differs from (94) by adding the mass term $
\frac{1}{2} \om^{2} ( x_{1}^{2} + x_{2}^{2}) $ , can be treated along
with the same line with a proper redefinition of parameters.
The interest in the corresponding classical system has been recently
renewed in connection with a study of non-abelian collective excitations
of the quark-gluon plasma \cite{Bla}.)
\newline
To find the first step Lie generator $ w_{1}^{(1)} $, it is convenient
to look for a solution in a form analogous to (96) \cite{Rob} :
\beq
w_{1}^{(1)} = \sum_{n,m = 0} ( \bzo^{n} \bzt^{m} a_{nm}^{(1)} -
(a_{nm}^{(1)})^{ \ast} \zo^{n} \zt^{m} + \bzo^{n} c_{nm}^{(1)}
\zt^{m} - \zo^{n} (c_{nm}^{(1)})^{ \ast} \bzt^{m} )
\eeq
We further substitute (96),(99) into eq. (43) and equate the like powers
of the quantum operators. The comparison of the coefficients provides
\bea
a_{nm}^{(1)} &=& f_{nm}  [ n \omo + m \omt + \mu ( n N_{2} + m N_{1}
+ n m ) ]^{- 1}   \\
c_{nm}^{(1)} &=& g_{nm} [ n \omo - m \omt + \mu ( n N_{2} -
m N_{1} )]^{-1}
\eea
Note that the denominator in (101) never goes to zero. For the only
nonvanishing $ g_{22} = 1 $ we obtain
\beq
c_{22}^{(1)} = [  \lm + 2  \lm ( N_{2} - N_{1})]^{-1}
\eeq
which does not turn to infinity at any eigenvalues of the operators
$ N_{1} \; , \; N_{2} $. Using the machinery of the proceeding section,
it is not difficult to calculate the next term in the approximate
quantum Hamiltonian (again, we retain only the terms yielding non-zero
vacuum expectation values) :
\bea
H^{(2)} &=& \omo (N_{1} + N_{2} ) + \mu N_{1} N_{2} + \alpha
+ \frac{ \var^{2}}{2} [ \frac{ (N_{1}^{2} - N_{1}) ( N_{2}^{2} - N_{2})}{
\omo + \omt - 4 \mu + \mu ( N_{1} + N_{2} )}  \nonumber \\
        &-& \frac{4 + 6 (N_{1} + N_{2}) + 9 N_{1} N_{2} + 2 ( N_{1}^{2}
+ N_{2}^{2} ) + 3 N_{1} N_{2} ( N_{1} + N_{2} ) + N_{1}^{2} N_{2}^{2}}{
\omo + \omt + 2 \mu + \mu ( N_{1} + N_{2} ) }   \nonumber   \\
        &+& \frac{(N_{1}^{2} - N_{1})( N_{2}^{2} + 3 N_{2} +2)}{ \omo
- \omt + 2 \mu + \mu ( N_{2} - N_{1} )} - \frac{ (N_{2}^{2} - N_{2})
(N_{1}^{2} + 3 N_{1} + 2 )}{ \omo - \omt - 2 \mu + \mu ( N_{2} - N_{1})}]
 + O( \var^{3} )
\eea
One can check by the term-by-term inspection that zeros do not appear
in the denominators in eq. (103). It can even be further speculated,
similarly to the line of reasoning of Ref. \cite{Rob}, that small
denominators do not show up at any order of the iterative procedure.
For the ground state energy we obtain from (103)
\beq
E_{0} = \frac{ 1}{2} ( 1 + \frac{ \lm}{2} -
 \frac{ \lm^{2}}{4} \frac{1}{ 1 + \frac{5}{2}  \lm } )
   + O( \var^{3} ) \; \; ,
\eeq
or, setting $  \lm = \frac{1}{2} $ ,
\beq
E_{0} \simeq \frac{1}{2} ( 1 + \frac{1}{4} - \frac{1}{36} )
\simeq 0.611
\eeq
This number is to be compared with the result of the numerical
calculation \cite{Med} which reads $ E_{0} \simeq 0.590 $. One sees
that our accuracy is rather good, of order 3.5 \%.
\newline
Let us now turn to the study of the approximate wave functions (WF)
(here we restrict ourselves by the first order formulas). The
unperturbed Hamiltonian $ H_{0} $ (96) has the eigenfunctions
\beq
\phi_{s}^{(0)} = | n_{1} n_{2} \rangle  \; \; \; , \; \; \;
s \equiv (n_{1} , n_{2} )   \; \; ,
\eeq
or, in the coordinate space
\beq
\phi_{(n_{1},n_{2})}^{(0)} (x_{1},x_{2}) = \frac{ ( \omo \omt)^{1/4}}{
\sqrt{ \pi  2^{n_{1} + n_{2}} n_{1} ! n_{2} ! }}
e^{ - \frac{ \omo x_{1}^{2} + \omt x_{2}^{2}}{ 2 }} H_{n_{1}}
( x_{1} \sqrt{ \omo} ) H_{n_{2}} ( x_{2}
\sqrt{ \omt} )
\eeq
where $ H_{n} (z) $ stands for the Hermite polynomial. After a simple
algebra we obtain the first order WF
\bea
\Psi_{(n_{1},n_{2})}^{(1)}
&=& \phi_{(n_{1},n_{2})}^{(0)} +
\frac{1}{2} \var [ \frac{ \sqrt{n_{1}(n_{1} -1)n_{2} (n_{2} -1)}}{
\omo + \omt + \mu ( n_{1} + n_{2} - 2)} \phi_{(n_{1} -2,n_{2}-2)}^{(0)}
            \nonumber \\
&+& \frac{ \sqrt{n_{1} (n_{1} - 1) (n_{2}+1)(n_{2} +2)}}{ \omo - \omt
+ \mu ( n_{2} - n_{1} +2)} \phi_{(n_{1}-2,n_{2}+2)}^{(0)} \nonumber \\
&-& \frac{ \sqrt{(n_{1} +1)(n_{1}+2) n_{2}
(n_{2} -1)}}{ \omo - \omt + \mu (n_{2} -n_{1} -2)}
\phi_{(n_{1}+2,n_{2} -2)}^{(0)}  \nonumber \\
&-& \frac{ \sqrt{ (n_{1} +1)(n_{1} +2)
(n_{2} +1)(n_{2} +2)}}{ \omo  + \omt + \mu (n_{1} + n_{2} + 2)}
\phi_{(n_{1} +2,n_{2} +2)}^{(0)}       \\
&+& \frac{(\gamma +2 n_{2}) \sqrt{n_{1}(n_{1} -1)}}{ \omo + \mu n_{2}}
\phi_{(n_{1}-2, n_{2})}^{(0)}
- \frac{( \gamma +2n_{2}) \sqrt{(n_{1} +1)(n_{1}
+2)}}{ \omo + \mu n_{2}} \phi_{(n_{1}+2,n_{2})}^{(0)} \nonumber \\
&+& \frac{( \gamma +2n_{1}) \sqrt{n_{2}(n_{2} -1)}}{ \omt  + \mu n_{1}}
\phi_{(n_{1},n_{2}-2)}^{(0)} -
\frac{( \gamma +2 n_{1}) \sqrt{(n_{2}+1)(n_{2}+2)}}{ \omt + \mu n_{1}}
\phi_{(n_{1},n_{2}+2)}^{(0)} ]   \nonumber
\eea
The following from (108) asymmetry of the ground state WF (GSWF) is
evidently an artifact of the asymmetry chosen in constructing the
perturbation theory. As the GSWF must possess a highest possible
symmetry, the final expression must be symmetrized in respect to
the replacement $ ( N_{1} \leftrightarrow N_{2} ) $ :
\bea
\Psi_{0}^{(1)} &=& \frac{1}{2} [ e^{ - \frac{ \omo x_{1}^{2} +
 \omt x_{2}^{2}}{ 2 }} [ 1 - \frac{  \lm}{8} ( 2 \gamma
x_{1}^{2}+ 2 \gamma x_{2}^{2}         \nonumber \\
&+&  \frac{((1 +  \lm) x_{1}^{2} - 1)( x_{2}^{2} -1)}{ 1 +
 \frac{5}{2}  \lm} - 2 \gamma \frac{2 +  \lm}{1 +
\lm} ) ] + ( x_{1} \leftrightarrow x_{2} ) ]
\eea
The obtained formula demonstrates that our expansion is controllable
in a vicinity of the origin. With moving away from the origin,
the correction constitutes about 50 \% of the unperturbed value
at $ x_{1} \simeq 2.0 $ , if one moves along the lone $ x_{2} = 0 $ ,
or $ x_{1} \sim x_{2} \sim 3.0 $ for the radial escape. ( Note that
the validity of the ansatz for the wave function for small $ x, y $
in the form $ \exp \times polynomial $ can be checked by the direct
substitution of such an ansatz into eq. (110) ).
\newline
To get approximate WF's in a large fields region, we need another
approximation. To this end, we make use of an intrinsically true
idea that, when one coordinate is large, the variables in the
Schrodinger equation get adiabatically separated \cite{Med}.
Note, however, that the method used in Ref. \cite{Med} does not
allow one to estimate corrections to the adiabatic separability,
which is the question of great importance if we are going to match
approximately two different expansions.
\newline
Let us write down the Schrodinger equation (in what follows, we set
 $ x_{1} \equiv x \; , \; x_{2} \equiv y $ )
\beq
-\frac{1}{2} ( \frac{ \pr^{2}}{ \pr x^{2}} + \frac{ \pr^{2}}{ \pr
y^{2}} ) \Psi ( x,y) + \lm x^{2} y^{2} \Psi (x,y) = E \Psi (x,y)
\eeq
and make in the region $ x > 0 $ the point canonical transformation
to the new adiabatic variables ( we are indebted to L. Franktfurt
for this suggestion) :
\bea
x & \rightarrow & u = x      \nonumber \\
y & \rightarrow & \tau = \sqrt{x} y
\eea
The Schrodinger equation then transforms into
\beq
\frac{ 1}{2} [ \frac{ \partial^{2}}{ \partial u^{2}}
 + u ( \frac{ \partial^{2}}{ \partial \tau^{2}}
 - \frac{2 \lm}{ \hbar^{2}} \tau^{2} )]  \Phi (u , \tau)
+ E \Phi ( u ,\tau )   =  ( \frac{ 1}{8} \frac{ \tau}{ u^{2}}
\frac{ \partial}{ \partial \tau}
- \frac{ 1}{2} \frac{ \tau}{u} \frac{ \partial^{2}}{ \partial
 u \; \partial \tau} ) \Phi ( u , \tau )
\eeq
where $ \Phi ( u , \tau) = u^{ - \frac{1}{4}} \Psi ( u ,
\tau u^{- \frac{1}{2}} ) $ , the multiplier is due to the Jacobian of the
transformation (111).
Note that in eq. (112) the variables get adiabatically
separated in the region of very large $ u $ provided
\bea
\left|  \frac{1}{4} \frac{ \tau}{ u^{2}} \frac{ \partial
\Phi}{ \partial
\tau} \right|  & \ll & \left| u \frac{ \partial^{2}
\Phi}{ \partial \tau^{2}}
\right|      \nonumber \\
\left| \frac{ \tau}{u} \frac{ \partial^{2}
\Phi}{ \partial  u \; \partial  \tau}
\right|        & \ll & \left| u \frac{ \partial^{2} \Phi
}{ \partial \tau^{2}} \right|
\eea
In the asymptotic region we neglect in the zero approximation the r.h.s.
and look for a solution in the form
\beq
\Phi_{0} = \sum_{n = 0}^{ \infty} \xi_{n} ( \tau) \eta_{n} (u)
\eeq
where $ \xi_{n} ( \tau ) $ satisfies
\beq
( \frac{ 1}{2} \frac{ \partial^{2}
}{ \partial \tau^{2}} - \lm \tau^{2} )
\xi_{n} ( \tau) = - \var_{n} \xi_{n}  ( \tau) \; \; ,
\eeq
i.e. the $ \tau $ - motion is the usual harmonic oscillator. The
resulting equation for $ \eta_{n} ( u) $ is
\beq
\frac{ d^{2} \eta}{ d u^{2}} - (2n + 1) \om  u
\eta +  2 E  \eta = 0
\eeq
where $ \om = \sqrt{2 \lm} $ .
The only physically accepted solution (damped as $ u \rightarrow
\infty $) is given in terms of the Airy function \cite{Lan} :
\beq
\eta_{n} (u) \; = \; const \; Ai \left[ \left( (2n +1) \om
 \right)^{1/3} ( u - \frac{ 2 E}{ (2n +1)  \om} ) \right]
\eeq
However, we are only allowed to retain the asymptotics of this formula,
thus
\beq
\Phi_{0} ( u , \tau) \simeq \sum_{n=0} A_{n} e^{ - \frac{ \om}{ 2
}} \tau^{2} H_{n} ( \tau \sqrt{  \om} )
\frac{1}{u^{ \frac{1}{4}}} e^{ - \frac{2}{3} \alpha_{n} ( u -
\beta_{n})^{3/2}} \; , \; u \rightarrow \infty
\eeq
where $ A_{n} $ are the normalization coefficients and
\beq
\alpha_{n} = \sqrt{ (2n+1) \om} \; \; , \; \;
\beta_{n} = \frac{2 E}{(2n+1)  \om}
\eeq
Let us now come back to eq. (112) and try to calculate corrections.
There is no room in (112) for the standard Green functions method
because of the non-separability of the variables at arbitrary
$ u , \tau $. To proceed, we substitute into (112) the ansatz
\beq
\Phi ( u , \tau) = \Phi_{0} (u, \tau) \Omega (u, \tau)
\eeq
and try to find the form of $ \Omega ( u , \tau) $ in a vicinity
of $ u \rightarrow \infty $. The boundary conditions on
$ \Omega ( u, \tau) $ are
\bea
\Omega(u ,\tau) \rightarrow 1 \; \; , \; \; u \rightarrow \infty \;
\; , \; \; \tau \; \; fixed  \\
| \Omega(u, \tau) | < e^{  \frac{ \om}{2 } \tau^{2}} \; \; ,
\; \; \tau \rightarrow \infty \; \; , \; \; u \; \; fixed
\eea
The equation for $ \Omega(u , \tau) $ reads (in what follows we denote
$ \Phi = \Phi_{0} \; \; , \; \; \Phi_{ \tau} = \frac{ \partial \Phi}{
\partial \tau} $, etc.)
\bea
( \frac{ \tau}{u} \Phi_{u \tau} - \frac{1}{4} \frac{\tau}{u^{2}}
\Phi_{ \tau} ) \Omega + ( 2 \Phi_{u} + \frac{ \tau}{u} \Phi_{ \tau} )
\Omega_{u}  + \frac{ \tau}{u} \Phi \Omega_{u \tau}   \nonumber \\
+ ( 2 \Phi_{ \tau} u + \frac{ \tau}{u} \Phi_{u} - \frac{1}{4}
\frac{ \tau}{u^{2}} \Phi ) \Omega_{ \tau} + \Phi \Omega_{uu}
+ \Phi u \Omega_{ \tau \tau} = 0
\eea
According to the boundary conditions, we look for a solution in
the form
\beq
\Omega( u ,\tau) = 1 + u^{-1/2} f_{1} ( \tau) + u^{-1} f_{2} ( \tau)
+ u^{-3/2} f_{3} ( \tau) + \ldots
\eeq
We further substitute this expansion into (122) and equate the like
powers of $ u $. The resulting equations for $ f_{1} \; , \; f_{2}
\; , f_{3} $
read
\bea
2 \xi'_{n} ( \tau) f'_{1} ( \tau) +
\xi_{n} ( \tau) f''_{2} ( \tau ) = 0 \\
2 \xi'_{n} ( \tau) f'_{2} ( \tau) +
\xi_{n} ( \tau) f''_{2} ( \tau ) = 0  \\
- \alpha_{n} \tau \xi'_{n}(\tau) + 2 \xi'_{n}( \tau) f'_{3}( \tau) +
\xi_{n} ( \tau) f''_{3} ( \tau) = 0
\eea
Let us first concentrate on the
equations (125), (126). It is easy to see that the non-trivial $ \tau $ -
-dependent solutions are excluded due to the boundary condition (122).
However, arbitrary constants are not fixed by the boundary condition
at the infinity. ( The acceptable solution for eq. (127) is a bit more
complicated: $ f_{3} ( \tau) = C_{3} + \alpha_{n} \int_{0}^{ \tau} dt
\xi_{n}^{-2} (t) \int_{0}^{t} dz z \xi'_{n} (z) \xi_{n} (z) $ ).
This means that our method is only giving a large-u
parametrization of the wave function. Recalling the Jacobian and
returning to the original variables, we finally arrive at
 the GSWF of the form (we retain here only the $ n=0 $ term from (118) )
\beq
\Psi ( x,y) = A e^{ -\frac{1}{2} x y^{2} - \frac{2}{3} ( x - \beta_{0}
)^{3/2} } \; ( 1 + \frac{C_{1}}{x^{1/2}} + \frac{C_{2}}{x}
 + \frac{ \Pi( \tau , C_{3} )}{ x^{3/2}} + \ldots )
\eeq
where $ A , C_{1} , C_{2} , C_{3}  $ are some constants and $ \Pi
( \tau , C_{3}) = C_{3} + \frac{1}{2} \int^{\tau} dt e^{
\alpha_{0}^{2} t^{2}} \Gamma ( \frac{3}{2} , \alpha_{0}^{2} t^{2} ) $,
$ \Gamma ( \gamma , z) $ stands for the incomplete Gamma function
(we will omit this term in what follows) .
Note that
the obtained WF is in no sense an analytic continuation of (109).
This is in the drastic contrast with one-dimensional quantum mechanics
where the mathematically rigorous matching of the WKB wave functions can
be given \cite{Lan}. We would like to mention that the appearance of
arbitrary constants in asymptotic large distance solutions is well
known in hydrodynamics \cite{Chang}. It is also known that small- and
large-distance expansions can be analytically matched only if the
small distance expansion has the infinite radius of convergence
\cite{Van}, which is not the case in our problem. Therefore, one has to
rely on an approximate (or numerical) matching. In this respect,
the situation is reminiscent of what one
encounters in the QCD sum rules method (provided we discuss the matching
problem for wave functions and not for Green functions).
There two
different asymptotic expansions, each of them being defined in its own
range, are numerically
matched in some intermediate region \cite{Shif}. It may be useful to
continue a bit this formal analogy. Our asymptotic adiabatic expansion
is a method of parametrization of the large distance (fields) behavior and
thus is analogous to the expansion over hadron states \cite{Shif}, the
parameters $ A \; , \; C_{1} \; , \; C_{2} $ being the "hadron masses".
The non-perturbative Kolmogorov technique substitutes the operator
product expansion at small distances. Probably a bit deeper  analogy
with quantum chromodynamics lies in the fact that the non-integrability
of the system (94) blocks an explicit construction of "quasiparticle"
weekly interacting operators in the intermediate region. It is not yet,
however, clear whether the non-integrability or just the mathematical
complexity of real QCD prevents one building hadron operators in terms
of the fundamental quark and gluon fields.
\newline
Inspired by this analogy, we have tried to find the coefficients
$ A , C_{1} , C_{2} $  by the numerical matching of the form
(128) with the WF (109) at moderate $ x \sim 1 - 2 $ in the vicinity
of the line $ y = 0 $ . Though there exists a variety of matchings
in three dimensions, the simplest method one can suggest is
a minimization of the difference between
the two wave function over the usual normalization measure with
the integration done over the "stability slice ", i.e. a continuous
limit of the usual least square method. As such procedure
can yields jumps of the derivatives, it must actually be supplemented
by some smoothing operation on the obtained functions. Within the
least square method, this problem is just ignored, which is
partially justified by the fact that we make the matching in the
exponentially suppressed region where derivatives are of order  of
the functions themselves.
\newline
We have found that there exists the "stability plateau"  ranged from
$ x \simeq 1.5 $ to $ x \simeq 2.5 $. With the matching in the intervals
of $ x = [ 1.5 , 2 ] $ or $ x = [2 , 2.5] $
  , the coefficients $ A , C_{1} , C_{2} $ change not more than by
10 \% . The found coefficients are
\beq
A  = 2.50 \pm 0.15   \; , \;
C_{1} = - 1.55 \pm 0.10 \; , \; C_{2} =  0.90 \pm 0.10
\eeq
with the measure parameter for the interval $ x = [1.5 , 2] $
\beq
M \equiv \frac{ \int_{1.5}^{2.0} dx \int dy ( \Delta \Psi)^{2}}{
\int_{1.5}^{2.0} dx \int dy \Psi^{2}} \simeq    11 \%
\eeq
where $ \Delta \Psi $ stands for the difference of the WF's (109)
and (128).
 The measure parameter over  the interval $  x = [2 , 2.5] $ is
a bit worse, of order 13 \% .
The main defect of the above matching is the fact that it deals solely
with the adjustment of the wave function without imposing any condition
on the derivatives. As the wave functions are known only approximately,
one cannot impose the condition of the continuity on the derivatives of
the wave functions. However, in view of the approximate character of
our solutions, in the intermediate region
it seems reasonable to proceed to the description of the wave functions
in terms of the distributions rather than the usual functions.
This means that we introduce some smoothing which characterizes the
actual accuracy of our calculations. Instead of the function at a
point $ x_{0} $, we get the "function" at some small region $ \Omega $
containing the point $ x_{0} $ :
\beq
\tilde{f}( \Omega) = \int_{x_{0} \in \Omega} f(x) \phi(x) d x  \; ,
\eeq
where $ \phi(x) $ is a differentiable function which is very strongly
peaked at $ x_{0} $ and vanishes outside the region $ \Omega $.
Then a precise meaning to the notion of matching two distributions
$ f_{1} $ and $ f_{2} $ can be given. The distributions $ f_{1} , f_{2} $
coincide in a open set $ \Omega $ when
\beq
\int_{ \Omega} f_{1} \phi = \int_{ \Omega} f_{2} \phi
\eeq
for every infinitely differentiable function $ \phi $ whose support
lies in $ \Omega $ \cite{Ric}. The analogous conditions hold for the
derivatives of $ f_{1} , f_{2} $ : $ \langle f_{1} , \phi' \rangle =
\langle f_{2} , \phi' \rangle $ .
For the purpose of the numerical
matching, the test functions $ \phi $ should be strongly peaked
at the stability plateau $ x = [1.5 , 2.5] $ whose existence is
indicated by the least square matching. In our opinion, such matching
in the distribution sense is adequate to the approximate character
of both solutions. Note that a more ambitious way of solving the
Schrodinger equation in the intermediate region would be a search
for a test function $ \phi $ satisfying the differential equation,
i.e. a solution in the form of the integral transformation.
\newline
For the approximate matching in the distribution sense, we have chosen
the test function in the form of the smoothed $ \delta $ function
\beq
\phi(x,y) = \frac{n m}{\pi} \exp [ - n^{2} ( x - x_{0} )^{2} - m^{2}
y^{2} ]
\eeq
and varied the parameters $ n , m  , x_{0} $ in the vicinity of the
values $ n = m = 1 $ and $ x_{0} = 2 $. Then the parameters $ A , C_{1}
, C_{2} $ have been determined from the matching of the integrated
wave functions and the their partial derivatives  over $ x $ ( the
matching of the partial derivatives over $ y $ is the triviality
when one chooses the test function (133) and neglects the term
$ \Pi ( \tau , C_{3} ) $ in (128) ). As in this case there are only
two equations for three unknown parameters, we have taken one of them
as the known one from (129) and found the other ones. We have found
the same values of the parameters with the accuracy quoted in (129).
We believe that the consistency between the two methods of the
matching confirms the reliability of the results (129). Still, as
we have seen, these two methods are not quite independent.
Another method of a matching could be a search for global conservation
laws, that would allow one to find the parameters $ A , C_{1} , C_{2} $
or some relations among them. Unfortunately, we have not succeeded
in this attempt.

\section{Concluding Remarks}

In this paper we have studied a quantum extension of the
Kolmogorov iterative technique. We have found  explicit solutions to
functional equations determining the high orders Lie generators and
Hamiltonians in both the classical and quantum cases and tested the whole
approach on a few problems. For YM quantum mechanics, we saw that the
Kolmogorov technique for the small distance regime must be supplemented by
the adiabatic large distance approximation, thus the matching problem arises.
 As the whole affair is still in
its infancy, one can hardly say beforehand whether it is able to suggest
a reliable calculation scheme for more realistic problems of quantum
mechanics
and quantum field theory. Nevertheless, as we have seen, the KAM
technique possesses a few very attractive features. The success in the
construction of the convergent iterative scheme for the quantum anharmonic
oscillator may allow one to hope that this technique could also be helpful
in more complicated problems, provided the method is easily generalizable
to arbitrary dimensional cases. The hope to overcome the Dyson instability
is due to the non-linear character of the KAM technique.
Moreover, because of its recurrent definition, this method seems to be
ideally suited for an analytical computerization. The need in the computer's
help is seen already at first Kolmogorov's steps. We want to emphasize,
however, that these difficulties are only technical and may be easily
handled. On the other  hand, the KAM technique is essentially
non-perturbative and suggests the iterative non-linear extension of the
Bogolyubov transformations method, when applied to infinite dimensional
systems. Thus, it can be of some interest for clarifying a connection
between perturbative and non-perturbative effects in quantum field theory.

The proposed technique could be further developed in a few possible
directions. One of them is working out a theory of non-linear canonical
transformations \cite{And}, which would allow one to calculate approximate
wave functions directly from the "non-coordinate" Lie transforms formalism.
Another problem is an improvement of the matching procedure and the use
of a more accurate smoothing technique analogous to that used in
classical mechanics. Also a search for global conservation laws seems to
be rather important for the matching problem. The matching procedure done
in this paper is very crude and, in particular, leads to an admixture of
excited states to the ground state wave function. Surely enough, the
matching must be improved to make this method work. Such an improvement
would be especially important for the matching large- and small- distance
solutions for highly excited states, where chaotic properties are expected
to manifest themselves in full. It would also be rather interesting to
reformulate the present approach in the path integral formalism and apply
it for a calculation of Green functions. Hopefully, this could be done in
the coherent states representation of the path integral. In such a form, this
method could be directly confronted with the Operator Product Expansion,
that might lead to a deeper understanding the phenomenological success of
the QCD sum rules method \cite{Shif}. As for the problem of calculation
of the spectrum, the generalization of the present approach is quite
straightforward. In particular, we hope that this method may be extended
to calculate the mass spectrum in the quantum YM theory with periodic
boundary conditions \cite{Lus}.
 Another possible direction
is a clarification of a connection of the present approach
with the quasiclassical quantization of periodic orbits which
hopefully could be useful for the quantum chaos studies.
In any case, a lot of work is needed before the suggested
method proves or disproves its power.

  I am very grateful to L. Franktfurt for numerous suggestions and
  valuable discussions.

 {\it Note added} When this paper has been completed, I became aware of
 the recent paper \cite{Sch} where similar ideas were developed and
 applied to the quantum anharmonic oscillator.



\end{document}